\documentclass{bmcart}

\PassOptionsToPackage{svgnames}{xcolor}
\usepackage[utf8]{inputenc}
\usepackage{graphicx}
\usepackage{etoolbox}
\usepackage{amssymb}
\usepackage{amsmath}
\usepackage{braket}
\usepackage{xspace}
\usepackage{placeins}
\usepackage{csquotes}
\usepackage{tabularx}
\usepackage{comment}
\usepackage{csquotes}
\usepackage{ifthen}
\usepackage[expansion,protrusion]{microtype}
\usepackage[colorlinks=true, urlcolor=teal,
            linkcolor=teal, citecolor=teal]{hyperref}
\usepackage{yquant}
\usepackage{adjustbox}
\usepackage[most]{tcolorbox}
\usepackage{tikz}
\usetikzlibrary{arrows.meta,calc, babel, positioning,shadows.blur,decorations.pathmorphing, decorations.pathreplacing, decorations.shapes}
\usepackage{isotope}
\usepackage{subfig}
\usepackage{siunitx}
\ifdefined\unit\else
  \ifdefined\NewCommandCopy
    \NewCommandCopy\unit\si
  \else
    \NewDocumentCommand\unit{O{}m}{\si[#1]{#2}}
  \fi
\fi
\newcommand{\eg}{\emph{e.g.\xspace}}

\startlocaldefs
\newif\ifdraft
\drafttrue
\ifdraft
\newcommand{\note}[1]{ {\textcolor{orange} { **: #1 }}}
\newcommand{\jknote}[1]{ {\textcolor{blue} { ***Johannes: #1 }}}
\newcommand{\grnote}[1]{ {\textcolor{cyan} { ***Georg: #1 }}}
\newcommand{\slnote}[1]{ {\textcolor{red} { ***Sebastian: #1 }}}
\newcommand{\tenote}[1]{ {\textcolor{purple} { ***ThEh: #1 }}}

\newcommand{\fdnote}[1]{ {\textcolor{green!50!black} { ***FD: #1 }}}
\newcommand{\kwnote}[1]{ {\textcolor{orange} { ***Karen: #1 }}}
\newcommand{\mynote}[1]{ {\textcolor{magenta} { ***MY: #1 }}}

\else
\newcommand{\note}[1]{}
\newcommand{\slnote}[1]{}
\newcommand{\tenote}[1]{}
\newcommand{\jknote}[1]{}
\newcommand{\grnote}[1]{}
\newcommand{\fdnote}[1]{}
\newcommand{\mynote}[1]{}
\newcommand{\kwnote}[1]{}
\fi

\endlocaldefs

\begin{document}
\begin{frontmatter}
\begin{fmbox}
\dochead{Research}
\title{Neutral Atom Quantum Computing Hardware: Performance and End-User Perspective}

%%%%%%%%%%%%%%%%%%%%%%%%%%%%%%%%%%%%%%%%%%%%%%
%%                                          %%
%% Enter the authors here                   %%
%%                                          %%
%% Specify information, if available,       %%
%% in the form:                             %%
%%   <key>={<id1>,<id2>}                    %%
%%   <key>=                                 %%
%% Comment or delete the keys which are     %%
%% not used. Repeat \author command as much %%
%% as required.                             %%
%%                                          %%
%%%%%%%%%%%%%%%%%%%%%%%%%%%%%%%%%%%%%%%%%%%%%%

%% example from template \author[
%   addressref={aff1},                   % id's of addresses, e.g. {aff1,aff2}
%   corref={aff1},                       % id of corresponding address, if any
%   noteref={n1},                        % id's of article notes, if any
%   email={jane.e.doe@cambridge.co.uk}   % email address
%]{\inits{JE}\fnm{Jane E} \snm{Doe}}
%\author[
%   addressref={qutac},
%   email={info@qutac.de}
%]{\inits{QUTAC}\fnm{} \snm{Quantum Technology and Application Consortium – QUTAC}}
\author[
   addressref={qutac,siemens},
   email={karen.wintersperger@siemens.com}
]{\inits{KW}\fnm{Karen} \snm{Wintersperger}}
\author[
   addressref={qutac,trumpf},
   email={florian.dommert@trumpf.com}
]{\inits{FD}\fnm{Florian} \snm{Dommert}}
\author[
   addressref={qutac,merck},
   email={thomas.ehmer@merckgroup.com}
]{\inits{TE}\fnm{Thomas} \snm{Ehmer}}
\author[
   addressref={qutac,sap},
   email={andrey.hoursanov@sap.com}
]{\inits{AH}\fnm{Andrey} \snm{Hoursanov}}
\author[
   addressref={qutac,bmw},
   email={Johannes.Klepsch@bmw.de}
]{\inits{JK}\fnm{Johannes} \snm{Klepsch}}
\author[
   addressref={qutac,oth,siemens},
   email={wolfgang.mauerer@siemens.com}
]{\inits{WM}\fnm{Wolfgang} \snm{Mauerer}}
\author[
   addressref={qutac,lhind},
   email={georg.reuber@lhind.dlh.de}
]{\inits{GR}\fnm{Georg} \snm{Reuber}}
\author[
   addressref={qutac,bosch},
   email={Thomas.Strohm@de.bosch.com}
]{\inits{TS}\fnm{Thomas} \snm{Strohm}}
\author[
   addressref={qutac,telekom},
   email={Ming.Yin@telekom.de}
]{\inits{MY}\fnm{Ming} \snm{Yin}}
\author[
   addressref={qutac,infineon},
   email={sebastian.luber@infineon.com}
]{\inits{SL}\fnm{Sebastian} \snm{Luber}}

%%%%%%%%%%%%%%%%%%%%%%%%%%%%%%%%%%%%%%%%%%%%%%
%%                                          %%
%% Enter the authors' addresses here        %%
%%                                          %%
%% Repeat \address commands as much as      %%
%% required.                                %%
%%                                          %%
%%%%%%%%%%%%%%%%%%%%%%%%%%%%%%%%%%%%%%%%%%%%%%

\address[id=qutac]{%                           % unique id
  \orgname{QUTAC}, % university, etc
  %\street{},                     %
  %\postcode{}                                % post or zip code
  %\city{London},                              % city
  \cny{Germany}                                    % country
}
\address[id=bmw]{%
  \orgname{BMW AG},
  \street{Bremer Str. 6},
  \postcode{80807}
  \city{Munich},
  \cny{Germany}
}
\address[id=bosch]{%
  \orgname{Robert Bosch GmbH, Corporate Research},
  \street{Robert-Bosch-Campus 1},
  \postcode{71272}
  \city{Renningen},
  \cny{Germany}
}
\address[id=lhind]{%
  \orgname{Lufthansa Industry Solutions GmbH \& Co. KG},
  \street{Sch\"utzenwall 1},
  \postcode{22844}
  \city{Norderstedt},
  \cny{Germany}
}
\address[id=merck]{%
  \orgname{Merck KGaA},
  \street{Frankfurterstr. 250},
  \postcode{64293}
  \city{Darmstadt},
  \cny{Germany}
}
\address[id=siemens]{%
  \orgname{Siemens AG},
  \street{Otto-Hahn-Ring 6},
  \postcode{81739}
  \city{München},
  \cny{Germany}
}
\address[id=sap]{%
  \orgname{SAP SE},
  \street{Dietmar-Hopp-Allee 16},
  \postcode{69190}
  \city{Walldorf},
  \cny{Germany}
}
\address[id=telekom]{%
  \orgname{Deutsche Telekom},
  \street{Winterfeldtstraße 21},
  \postcode{ 10781},  
  \city{Berlin},
  \cny{Germany}
}
\address[id=trumpf]{%
  \orgname{TRUMPF SE + Co. KG},
  \street{Johann-Maus-Straße 2},
  \postcode{71254}
  \city{Ditzingen},
  \cny{Germany}
}
\address[id=infineon]{%
  \orgname{Infineon Technologies AG},
  \street{Am Campeon 1-15},
  \postcode{85579}
  \city{Neubiberg},
  \cny{Germany}
}
\address[id=oth]{%
  \orgname{Technical University of Applied Sciences Regensburg},
  \street{Galgenbergstraße 32},
  \postcode{93053}
  \city{Regensburg},
  \cny{Germany}
}

%%%%%%%%%%%%%%%%%%%%%%%%%%%%%%%%%%%%%%%%%%%%%%
%%                                          %%
%% Enter short notes here                   %%
%%                                          %%
%% Short notes will be after addresses      %%
%% on first page.                           %%
%%                                          %%
%%%%%%%%%%%%%%%%%%%%%%%%%%%%%%%%%%%%%%%%%%%%%%

\begin{artnotes}
%\note{Sample of title note}     % note to the article
%\note[id=n1]{Equal contributor} % note, connected to author
\end{artnotes}

\end{fmbox}% comment this for two column layout

\begin{abstractbox}

\begin{abstract} 
We present an industrial end-user perspective on the current state of quantum computing hardware for one specific technological approach, the neutral atom platform. Our aim is to assist developers in understanding the impact of the specific properties of these devices on the effectiveness of algorithm execution. Based on discussions with different vendors and recent literature, we discuss the performance data of the neutral atom platform. Specifically, we focus on the physical qubit architecture, which affects state preparation, qubit-to-qubit connectivity, gate fidelities, native gate instruction set, and individual qubit stability. These factors determine both the quantum-part execution time and the end-to-end wall clock time relevant for end-users, but also the ability to perform fault-tolerant quantum computation in the future. 
We end with an overview of which applications have been shown to be well suited for the peculiar properties of neutral atom-based quantum computers.
\end{abstract}

\begin{keyword}
\kwd{Neutral atom quantum computers}
\kwd{Review}
\kwd{Quantum computing platforms}
\kwd{Performance metrics}
\kwd{Benchmarks}
\end{keyword}

\end{abstractbox}

\end{frontmatter}

\section{Introduction}

\subsection{Quantum computing hardware platforms}

Quantum computers have been a topic of intense research and development, with the promise to revolutionise the field of computing. The primary motivation for exploring the application of quantum computers lies in their potential to solve certain types of problems better, faster or to tackle problems which are intractable for classical computers.

Classical computers are limited to performing computations in a sequential and deterministic manner. Quantum computers, however, can exploit peculiar features of quantum physics, which could enable super-linear or even exponential speedups for specific problems. However, the development of quantum computers and their applications is still in its early stages, and many technical challenges remain to be overcome. Moreover, not all problems will benefit from using quantum computers, and there are still many unanswered questions about the scalability of quantum hardware and the practicability of quantum algorithms. Nonetheless, the potential impact of quantum computing on areas such as cryptography, optimisation, drug discovery, and materials science is significant, making it an active area of research not only in academia but as well in industry.

There are various approaches to building a quantum computer~\cite{Preskill_2018}, each relying on different physical systems to create, connect, and manipulate qubits \cite{bsi_2020, fedorov2022quantum, ezratty_2021}. Some of the most promising approaches include: 

\begin{itemize}
\item Superconductors: This approach uses superconducting resonant circuits to build qubits. The circuits operate at very low temperatures, allow for fast gate operations, but currently still suffer from low coherence times and limited connectivity.

\item Ion traps: These use trapping techniques to trap ions in a vacuum chamber, laser cooling techniques to cool the ions, and electromagnetic pulses in the optical, microwave, or radio frequency range to manipulate their quantum states. This approach has been successfully used to manufacture quantum computers with high levels of coherence and very high connectivity. However, calculation speed and scaling to large numbers of qubits remain a challenge.

\item Neutral atoms: This approach uses laser cooling and trapping techniques for neutral atoms and manipulates their quantum states using optical or microwave pulses. It offers long coherence times, scaling in 2D or even 3D, and fair connectivities by long-range interactions (Rydberg states). The main challenges include further improvement of the two-qubit gate fidelities and gate operation speeds.

\item Photons: Photonic quantum computers use light to carry quantum information, which promises very good scalability. The virtually non-existent interaction between photons imposes a challenge to implementing gate operations. Thus, most often, so-called measurement-based quantum computing is being used.

\item Spins in semiconductors: In this approach, the spin of electrons or nuclei, typically in silicon, is used as the basis for qubits. It is compatible with existing semiconductor fabrication techniques, making it a promising option for scaling. However, the feasibility of such scaling concepts still has to be shown. 

\item NV centers in diamond: This approach uses nitrogen-vacancy (NV) centers in diamond to create qubits. The NV center is a defect that is usually manipulated using laser techniques and---at least theoretically---offers room-temperature operation. 

\end{itemize}

Each approach has its advantages and challenges, and research is ongoing to determine which approach is most suitable for scaling up quantum computers to a practical level. Furthermore, the specifics of each platform are expected to be more favourable for a limited and platform-specific set of algorithms, at least during the early phases of quantum computing.

\subsection{Benchmarks}

Industry users of quantum computers are mainly interested in two central questions:  (i) \textit{when will the hardware be available to run economically relevant applications that are faster than classical computers}, and (ii) \textit{which hardware is required for this?} These questions concern the performance of quantum computing hardware, and even for future fault-tolerant quantum computers, answering them will be difficult, as is the case for classical high-performance computing. In the current era of NISQ (noisy intermediate-scale quantum)-devices~\cite{Preskill_2018}, the challenge of answering these questions will be even greater, as highlighted in recent research~\cite{Li_2021}.

\newcommand{\LINPACK}{\textsc{LINPACK}\xspace}
To assess the performance of conventional computers, \emph{benchmarks} are being used, which implement particular \emph{metrics} to quantify the capabilities of computing hardware.  One of the most established benchmarks in conventional high-performance computing (HPC) is \LINPACK~\cite{Dongarra_etal_LINPACK_2002}. \LINPACK involves different numerical tasks, solved with the \LINPACK library, and results in a measure called FLOPS.  While \LINPACK may serve well to assess the performance of HPCs for numerical tasks, it is less suitable to evaluate an HPC for other classes of applications.

Quantum computers are in a nascent stage, which makes performance evaluation difficult. They follow a computing paradigm distinct from that of classical computers, which requires reproducible benchmarks~\cite{mauerer:22:q-saner} that address their unique properties and architecture layers.  The following paragraphs give an overview of established benchmarks for quantum computing that enable an evaluation of its computing capabilities.

One category of benchmarks focuses on fundamental physical functionalities, including qubit count, qubit stability, qubit coherence, and gate fidelity. %elaborated in section~\ref{subsec:explanation_criteria}.
Termed as \textit{low-level benchmarks}, they are crucial in developing and improving quantum hardware. However, predicting application performance directly from these metrics is challenging due to the varying influence of different parameters on the calculation outcome and their complex interplay.

Another benchmarking approach evaluates the quantum computer's overall characteristics. These benchmarks execute random unitaries~$U$ and compare the predicted and measured probability distributions of outcomes. An early instance is \emph{Google's cross entropy}~\cite{Boixo_etal_2018}, where random unitaries are generated using a highly generic quantum circuit with random parameter values. One can show that the probability distribution of outcomes for a random unitary in a perfect quantum computer follows the \emph{Porter-Thomas} distribution. However, due to the noise in real quantum computers, the measured distribution deviates from the Porter-Thomas distribution, with complete decoherence resulting in a uniform distribution. The cross-entropy fidelity~${\cal F}_\text{XEB}$ measures the deviation of the measured distribution from the Porter-Thomas distribution.

Cross~\etal~at IBM~\cite{Cross2019} extended the concept of cross-entropy fidelity to develop the \emph{quantum volume} metric. They introduce an additional step in which a circuit-to-circuit compiler searches for an optimal implementation of a selected random circuit for specific hardware. Thus, the benchmark outcome reflects the strengths of different hardware platforms. The quantum volume uses random quadratic circuits with the same circuit width (number of qubits) and depth. As the size of the random circuit increases, the measured distribution deviates more from the Porter-Thomas distribution. The quantum volume is the maximum size of a quadratic random circuit where this deviation remains below a specific limit. Although the quantum volume is useful for evaluating the overall performance of a quantum computer, it does not consider specific applications.

Lubinski~\etal~\cite{Lubinski_etal_2021,Lubinski_etal_2023} aim to evaluate the suitability of a quantum computer for running specific quantum algorithms by proposing \emph{algorithm-oriented benchmarks}, thereby coming closer to the goal of measuring the suitability of a quantum computer to run a specific application. Unlike random circuits, these benchmarks use specific quantum algorithms such as the Quantum Fourier Transformation or Grover's Search algorithm, which have different circuit depths for the same number of qubits (width). Therefore, in contrast to the quantum volume, these benchmarks require two numbers to specify the behaviour: width and depth of the circuits. To assess the performance of running a particular algorithm, instances of different width and depth of the algorithm are executed. While the largest instance that still yields a minimum fidelity characterises the performance of the quantum computer for the particular quantum algorithm, the algorithm-oriented benchmarks do not contain application-specific metrics - which depend on more factors, as detailed below. 

All metrics above only marginally look at the execution speed. \emph{Circuit layer operations per second} (CLOPS)~\cite{Wack_etal_CLOPS_2021} measures the number of layers of a parameterised model circuit that can be executed per unit of time.  It is important to stress that various parts of the quantum hardware-software stack contribute to CLOPS, including the repetition rate of the quantum processor, the speed at which gates run, the runtime compilation, the amount of time it takes to generate the classical control instructions, and finally, the data transfer rate among all units.

Other protocols, which also aim at using specific quantum algorithms to evaluate the 
performance of quantum computers, are QASMBench~\cite{ang_qasm_benchmarking_2022}, Q-Score~\cite{qscore_2023}, QPack~\cite{koen_qaoa_benchmarking_2022}, Quantum chemistry benchmark~\cite{alexander_chemistry_benchmarking_2022}, QUARK~\cite{quark_2022} and others. Colin~\etal~\cite{Colin_standard_benchmarking_2022} propose an initial version of a Quantum benchmark suite with standardised key performance indicators (KPIs).

In the NISQ-era, a quantum computer's performance depends not only on the properties of the hardware being used, but also on the specific problem being solved. A particular type of quantum computer might suit a specific problem and algorithm.
Thus, application-level benchmarks are valuable, particularly for practitioners who must select the most suitable computing platform for their problems. However, as highlighted in Ref.~\cite{amico2023defining}, application-level benchmarks can be affected by \enquote{benchmark optimisations}, where hardware-specific pre- or post-processing steps are used to alter the general validity of benchmark runs. Moreover, quantum error mitigation techniques can impact the validity of benchmark runs since solution quality can be traded against time-to-solution. 
Application-level benchmarks should, therefore, not be considered in isolation---thorough descriptions of low-level properties of quantum computers should accompany them. Moreover, knowledge about the physical mechanisms that govern a quantum computing device and its sources of errors can help to develop hardware-efficient algorithms and schemes for quantum error correction that use fewer resources. Low-level characterisations enable to identify bottlenecks, provide insights into higher-abstraction benchmarks and give indications about future developments. 

This paper discusses the most relevant properties and current state-of-the-art for one quantum hardware platform, namely neutral atoms. Similar reviews of other quantum computing hardware platforms are planned for the future. The neutral atom platform is still relatively young, but has many promising properties, especially in view of industry-relevant use-cases such as optimisation problems. The remainder of the paper is structured as follows: The physical foundations underlying neutral atom quantum computers are explained in Section~\ref{sec:foundations}. The most relevant hardware properties are briefly described in Section~\ref{subsec:explanation_criteria}, where relevant metrics for neutral atom quantum computers are listed. Section~\ref{subsec:sig_applications} describes the significance of these hardware properties for applications, while Section~\ref{subsec:providers} gives a brief overview of providers of neutral atom quantum computers. The last section~\ref{sec:discussion} summarises the main advantages and disadvantages of neutral atoms quantum computers and discusses examples of applications that profit particularly from the characteristics of this hardware platform.

\section{Physical foundations of the neutral atom platform}~\label{sec:foundations}

To better understand the properties and performance of neutral atom quantum computers, we start with a brief overview of the fundamental physics and resulting implementations. This section assumes knowledge about some advanced physics topics, which not every reader might be familiar with. But skipping the section does not prevent the reader from understanding the remainder of the paper.

\subsection{Architecture}
\begin{figure}[htb]
\centering
\includegraphics[width=0.98\textwidth]{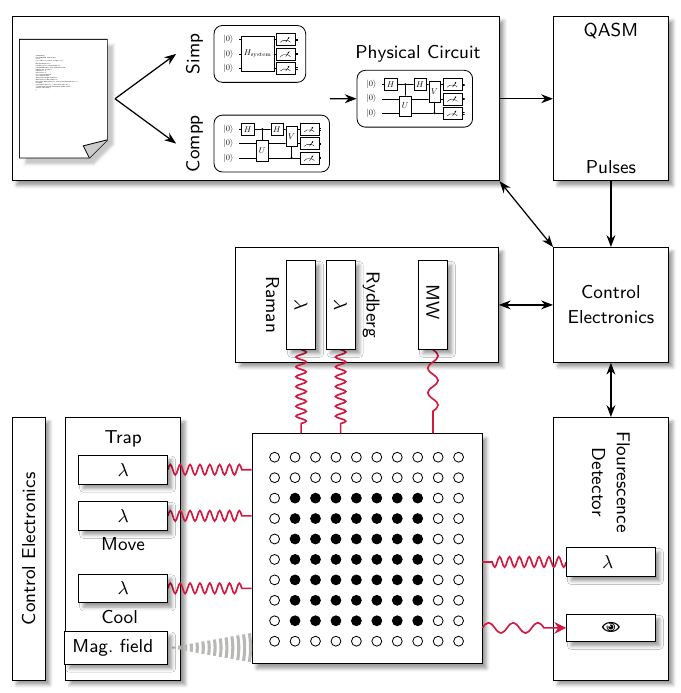}
    \caption{Schematic component overview of a cold atom quantum
      computer (the sequence of operations required to 
      perform an actual computation is shown in Fig.~\ref{fig:sequence}).}
    \label{fig:overview}
\end{figure}

Consider Figure~\ref{fig:overview}, which provides a schematic overview of
the main components that comprise a cold atom quantum computer. 
For now, we focus on their essential interplay, and discuss
relevant details in the following sections. 

An ultra-high vacuum cell that contains a geometric arrangement of neutral atoms
trapped by optical tweezers (and other mechanisms) lies at the core of the arrangement.
Before any computational steps can be performed on the qubits encoded in some of the
atoms' physical degrees of freedom, they need to be cooled down by a multi-stage
cooling process that especially involves stochastic Doppler cooling using a
laser. To place the atoms on fixed locations (and keep them in place) once they
have been sufficiently cooled down, a trap laser is employed, possibly augmented by means of moving the atoms around using another mobile trap. 
Orchestrating the involved lasers requires precise control at time scales that are
only accessible to special-purpose control electronics.

To issue computational steps by controlled manipulation of the trapped
atoms, laser- and sometimes also microwave pulses are employed. The required
pulse sequences are derived from instructions that need to be loaded prior
to execution into another set of classical high-speed electronic control elements.
The instructions are, in turn, derived from either a gate sequence (quantum
computing) or a more general Hamiltonian (quantum simulation) that has been
generated by an appropriate compiler from program source code, and may be implemented
in any of the many currently available quantum frameworks.
Any quantum computing sequence comprises a number of orchestrated
operations, which are discussed in the following sections and illustrated in Figure~\ref{fig:sequence}.

\subsection{Implementing qubits}\label{sec:implementing}
In this section, the physical implementation of qubits using cold atoms is explained, as well as the mechanisms to cool and trap atoms. The different steps necessary to perform a quantum computation are described in Subsection~\ref{sec:quantum_computing_sequence} and Figure~\ref{fig:sequence}.

\subsubsection{Qubit encoding}\label{sec:qubit_encoding}
In the cold-atom approach, a qubit is represented by a single atom. Most commonly, atoms from the first two groups of the periodic table are used, since these are well-suited for cooling and trapping due to their electronic structure. Prevalent examples include Rubidium (\isotope[87]{Rb})~\cite{henriet_quantum_2020, bluvstein_quantum_2022}, Caesium (\isotope[133]{Cs})~\cite{graham_multi-qubit_2022} and Strontium (\isotope[87]{Sr})~\cite{barnes_assembly_2022, park_cavity-enhanced_2022}. Moreover, the availability of lasers in the respective range of wavelengths plays an important role when choosing a certain kind of atom to build a quantum computer.  

\begin{figure}[htb]
    \centering
    \includegraphics[width=0.4\textwidth]{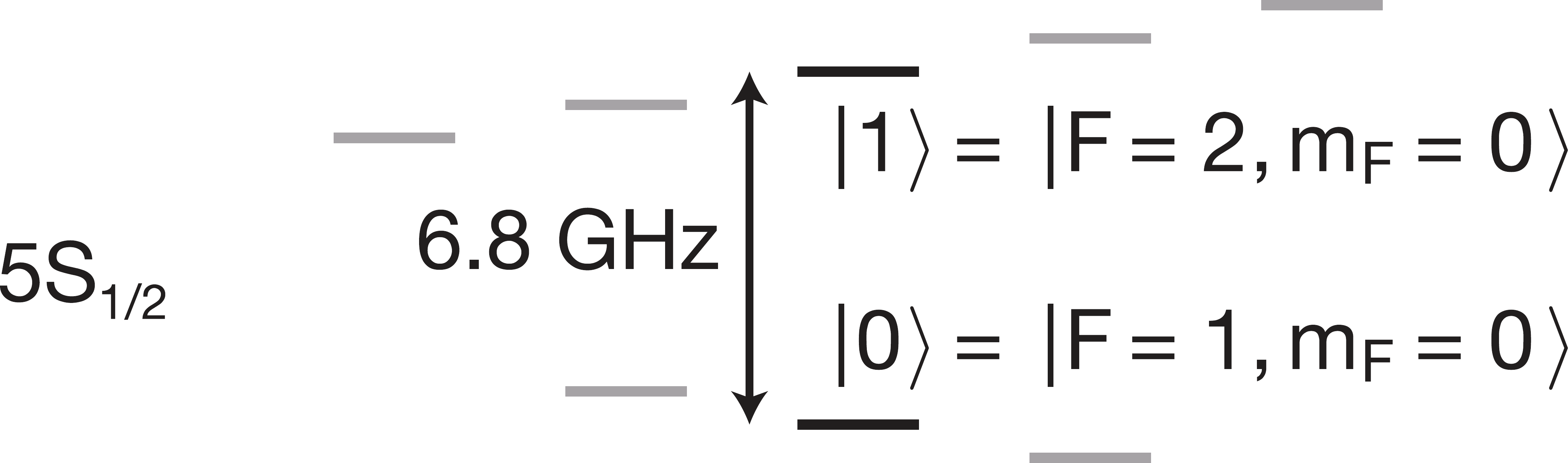}
    \caption{Ground state energy levels of \isotope[87]{Rb}. The two levels marked in black are being used as qubit states. Image adapted from~\cite{bluvstein_quantum_2022}.}
    \label{fig:level_scheme_Rb87}
\end{figure}

Depending on the choice of atom, the qubit states~$\ket{0}$ and~$\ket{1}$ are commonly realised via electronic spin states, e.g., for \isotope[87]{Rb}~\cite{henriet_quantum_2020, bluvstein_quantum_2022}, \isotope[133]{Cs}~\cite{graham_multi-qubit_2022}, or \isotope[87]{Sr}~\cite{park_cavity-enhanced_2022}. In Fig.~\ref{fig:level_scheme_Rb87}, the ground state energy levels of \isotope[87]{Rb} are shown schematically, which represent different possible energies of the single outermost electron. These arise due to the coupling of the spin of this single electron with its angular momentum and the spin of the atomic nucleus (being denoted by~$F$) and additional interaction with an external magnetic field (denoted by~$m_F$). The two lines marked in black illustrate a possible choice of the two qubit states. In this case, transitioning from~$\ket{0}$ to~$\ket{1}$ corresponds to flipping the electron spin relative to the nuclear spin. The energy difference between the two states can directly be converted to a frequency, which lies in the microwave range ($6.8\,\unit{GHz}$).

Using atoms with an energy level structure such as Strontium, also nuclear spin states can be employed as qubit states. This has the advantage of longer decoherence times, since nuclear spins interact much less with their environment than their electronic counterparts. On the other hand, the control of nuclear spin qubits is technically more challenging, involving additional magnetic fields and lasers to isolate the qubit transition~\cite{barnes_assembly_2022}.

\subsubsection{Cooling and trapping}\label{sec:trapping}

\begin{figure}[htbp]
\includegraphics[width=1.0\textwidth]{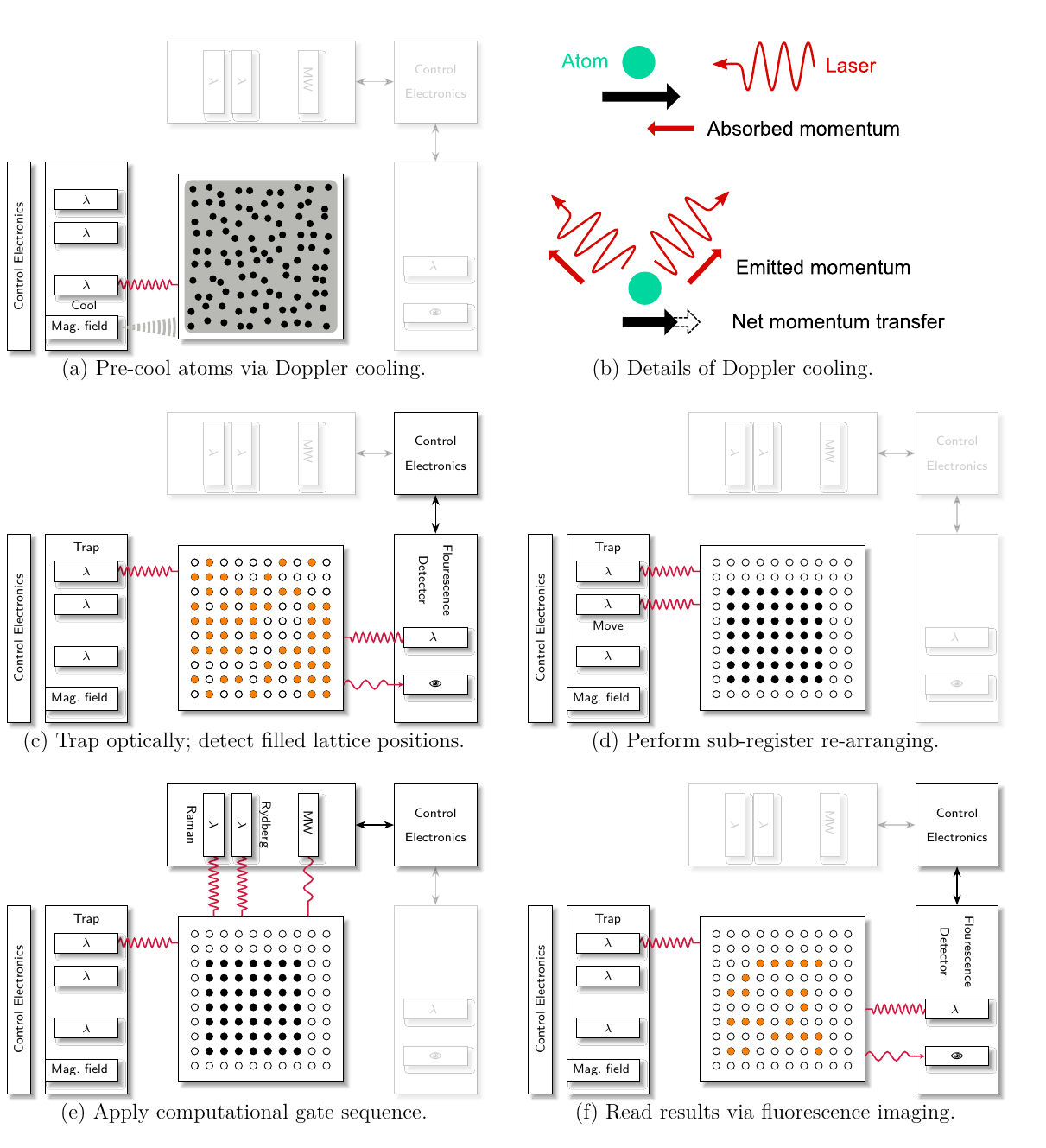}
    \caption{Initialisation and computation sequence.
    (a) First, a trap volume is filled with atoms captured
    in a magneto-optical trap, and pre-cooled using
    Doppler cooling (b): An atom absorbs momentum directed opposite to its velocity (top). Then, spontaneous emission of photons in an arbitrary directions leads to a net momentum transfer opposite the direction of travel of the atoms (bottom). The pre-cooled atoms are then trapped in a grid using optical technologies.
    Since the grid positions are filled with non-unit probability, fluorescence imaging is used to detect filled positions (c), the occupied positions are moved into adjacent positions (d). After applying
    the desired sequence of one- and two-qubit gates (e),
    the final qubit states are read using a collective
    imaging measurement (f).}
    \label{fig:sequence}
\end{figure}

To execute well-defined operations on neutral atom qubits, the atoms need to be cooled (slowed down) and trapped. Cold atom-based systems typically work at room temperature, but the atoms themselves are cooled to below~$1\,\unit{mK}$ using laser light.
In general, there are various different techniques to cool atoms by interaction with light. As a first preparation step before the trapping (see Fig.~\ref{fig:sequence}a), often so-called magneto-optical traps (MOTs)~\cite{dieckmann_2DMOT_1998} are being used which employ \emph{Doppler cooling}: a laser beam is directed onto a hot atom which moves at a certain velocity, as illustrated in Fig.~\ref{fig:sequence}b.
The frequency of the laser beam is adapted, such that it matches the resonance frequency of a certain atomic transition, taking into account the Doppler shift. By absorbing a photon from the laser beam, the atom does not only gain energy, but also momentum. Subsequently, the photon is being re-emitted by the atom, however, into an arbitrary direction. Scattering many photons per atom, this leads on average to a net momentum transfer opposite to the direction of travel and thus the atom is being slowed down. Since the laser frequency is adapted to match the Doppler-shifted transition of atoms moving towards the beam, the probability of absorption is much less for atoms travelling in the same direction as the light. 

To cool atoms travelling in arbitrary directions, one uses three pairs of laser beams that are directed towards the atom, in the direction of the six semi-axes of a three-dimensional Cartesian coordinate system.
A magnetic field is applied to prevent the atoms from drifting out of the cooling area and thereby creates not only a velocity-dependent, but also a space-dependent force. 

After pre-cooling the atoms, they are captured by optical traps, as illustrated in Fig.~\ref{fig:sequence}c. In general, optical trapping uses the so-called dipole force~\cite{grimm_dipole_2000}: the time-periodic electromagnetic field of the laser light induces a dipole moment in the atom, that is, a separation of positive and negative electric charges. This induced dipole moment interacts with the light field. Depending on the frequency difference between the two oscillating dipoles, the atom experiences either an attractive or repulsive force. Most commonly, the frequency of the light is smaller than that of the atomic transition (\emph{red-detuning}), which leads to an attractive force, trapping the atom near the focus of a laser beam.

For quantum computers based on neutral atoms, often arrays of so-called optical tweezers are being used, which can be realised using spatial light modulators (SLM). These produce an array of several tightly focused laser beams, each of which can trap a single atom. The trap arrays can be created in arbitrary geometric arrangements~\cite{nogrette_single-atom_2014}, in two and also three dimensions~\cite{barredo_synthetic_2018} and can prepare relatively large systems of hundreds or thousands of particles.
Typical distances between the individual traps are about~$3\,\unit{\micro m}$, which allows for addressing of single qubits with laser beams to execute gate operations. As the array of small traps needs to be created from a single laser beam, the number of traps that can be generated in this way is mainly determined by the available laser power, ensuring that each individual trap has a sufficient depth and the atoms cannot escape from the traps. The lifetime of an atom in such a microtrap is mostly limited by the residual vacuum pressure and typically lies in the range of~$10-20\,\unit{s}$~\cite{ebadi_quantum_2021, schymik_enhanced_2020}.

Another approach is to trap the atoms using a line grid array~\cite{graham_multi-qubit_2022} created by overlapping several line-shaped laser beams instead of the tweezer array, leading to comparable lifetimes of~$\sim 10\,\unit{s}$. Large-scale 3D multilayer configurations can also be achieved via a microlens-generated Talbot optical lattice~\cite{Schlosser_etal_2019}.

\subsubsection{Quantum computing sequence}\label{sec:quantum_computing_sequence}

After laser-cooling the optical trap is being switching on, leaving the atoms trapped with a probability of about~$50-60\,\unit{\percent}$ only~\cite{schymik_enhanced_2020}, whereas each trap contains one atom or no atom. To create a computational array with unit filling, the atoms are usually rearranged by additional mobile traps generated by acousto-optical deflectors (AODs). For this purpose, the atoms need to be imaged first: sending light with the appropriate frequency leads to fluorescence which is collected on a charge-coupled device (CCD) camera, as illustrated in Fig.~\ref{fig:sequence}c. In this way, all atoms are imaged simultaneously. On the resulting image, atoms are visible as bright spots and empty traps as dark spots. Knowing the initial arrangement of the atoms, these are then transported between the array traps using the mobile traps, as illustrated in Fig.~\ref{fig:sequence}d.

The resulting atomic array with unit filling constitutes the starting point for digital or analogue quantum computation and simulation. To perform digital quantum computing, single- and multi-qubit gates are executed in form of laser or microwave pulses (Fig.~\ref{fig:sequence}e), which is described in more detail in the next section. In the case of analogue operations, the desired Hamiltonian is switched on for a certain time usually also by applying laser pulses or additional magnetic fields. The final state of all qubits after the operations is read out simultaneously by fluorescence. A common way to distinguish between the two qubit states is to use light with an excitation energy that is sufficient to kick atoms in state~$\ket1$ out of the traps but not atoms in state~$\ket0$. Thus, the atoms in state~$\ket{1}$ appear as dark spots in the resulting image, while atoms in~$\ket{0}$ are visible as bright spots, as shown in Fig.~\ref{fig:sequence}f. Recently, also mid-circuit measurements have been demonstrated in a cold atom setup~\cite{graham_midcircuit_2023}.

To start a new computation run, the process described above needs to be traversed again starting from the initial cooling step.

\subsection{Implementing logic gates}\label{sec:gates}

So far we discussed how to initialise and measure the qubits of a cold-atom quantum computer. Next, we dive deeper into another crucial part of the computing sequence: the manipulation of the qubits.

\subsubsection{Single-qubit gates}\label{sec:sq-gates}

Quantum logic gates on neutral atom qubits are realised by applying laser and/or microwave pulses. In general, these driving fields, if being close to resonance, excite coherent oscillations between the qubit states~$\ket{0}$ and~$\ket{1}$. Applying the field for a certain time with a defined frequency, intensity and phase allows to create superpositions between the qubit states and thus to realise arbitrary rotations on the Bloch sphere. 

For the atomic species typically chosen in neutral atom quantum computers, the qubit transition lies in the microwave range (see Sec.~\ref{sec:qubit_encoding}). This transition could be driven directly by microwave pulses, but the atoms usually are so close to each other in the trap that a microwave pulse, due to its large wavelength, cannot be focused to a single atom. Instead, it is common to excite single qubit gates optically by driving \emph{Raman transitions}: using two laser beams with frequencies that differ by the value of the qubit transition frequency, the qubit levels are effectively coupled via excitation to a higher-lying state. The Raman beams are either directly focused onto one~\cite{henriet_quantum_2020} or several qubits or applied to a line of qubits simultaneously~\cite{levine_dispersive_2022}. These global Raman beams can be combined with additional addressing beams that shift the transition frequency of certain qubits, such that they are not resonant with the former anymore, to allow for local rotations. In a similar approach, global microwave pulses can be combined with addressing lasers that shift the qubit transition frequency in or out of resonance with the microwave~\cite{graham_multi-qubit_2022}. In general, to act on several selected qubits, the corresponding Raman or addressing beam needs to be divided and steered, which can be realised using AODs. With the protocols described above, it is possible to execute the same single-qubit gate on several qubits simultaneously. To execute different single-qubit gates on several qubits at the same time, multiple different beams are necessary, as their properties such as the pulse length, intensity or phase need to vary.

\subsubsection{Two-qubit gates}\label{sec:tq-gates}

The native gate set of cold-atom quantum computers usually consists of single-qubit rotations and the \textit{CZ}-gate, which can be used to create a \textit{CNOT}-gate by surrounding the target qubit with two Hadamard gates. To generate entanglement between two atoms, as required to implement a two-qubit gate, they need to interact with each other, such that the state of the target atom depends on the state of the control atom. Due to the relatively large distances between the atoms in the optical traps, they would normally not see each other. 

\begin{figure}[!htp]
    \centering
    \includegraphics[width=0.35\textwidth]{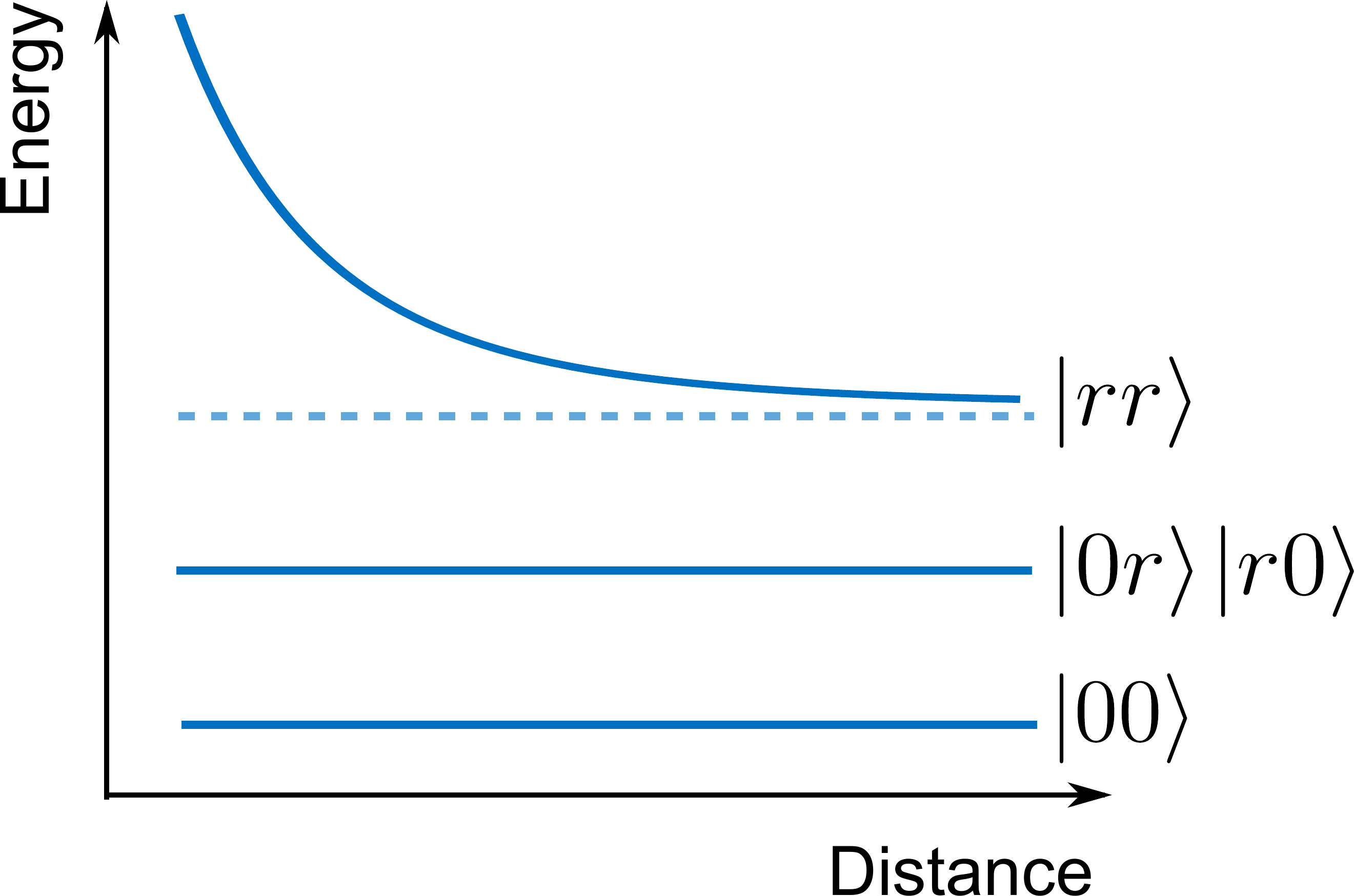}
    \caption{Illustration of Rydberg blockade: The energy of two atoms being excited to the Rydberg state~$\ket{r}$ diverges with their inverse distance.}
    \label{fig:ry_blockade}
\end{figure}

The necessary interaction is realised by exciting atoms to states with very high energy, called \emph{Rydberg} states~\cite{Isenhower2009, gaetan_observation_2009, wilk_entanglement_2010}. In a simplified picture, an atom in a Rydberg state has a large spatial extent, since the outermost electron (i.e., in the case of an alkali atom like Rubidium) can move far apart from the atomic core due to its high energy. That leads to a large dipole moment and a long-range, repulsive interaction between two Rydberg atoms. If two atoms are close to each other, only one of them can be excited to the Rydberg state, as the energy needed to excite both atoms diverges with the inverse distance between them, as illustrated in Fig.~\ref{fig:ry_blockade}. This \emph{Rydberg blockade} is used to create a state-dependent interaction between two qubits that can be switched on and off arbitrarily. The ability to introduce these strong and widely tunable interactions has led to a successful application in many-body physics and analogue quantum simulation~\cite{Morgado_Whitlock_2021}.

Since a high energy is needed to excite an atom to a Rydberg state, usually two lasers are being used to couple one of the qubit states and the Rydberg state via an intermediate energy level. To execute a two-qubit gate, these two beams are both divided and focused onto the two corresponding atoms simultaneously. Due to its high energy, an atom in a Rydberg state cannot be trapped anymore and, therefore, the trapping light is switched off during the application of the Rydberg pulses~\cite{levine_parallel_2019} (not shown in Fig.~\ref{fig:sequence}e). As the two-qubit gate sequence is fast compared to any residual motion of the cold atoms, the atoms are captured again when switching the traps back on afterwards. 

There are two common schemes to realise a \textit{CZ}-gate which use two~\cite{levine_parallel_2019} or three~\cite{henriet_quantum_2020} laser pulses, respectively. While the latter scheme is simpler, the former can be performed much faster within~$\sim 400\,\unit{ns}$ instead of~$\sim 1\,\unit{\micro s}$ for the three-pulse approach. In each case, the Rydberg state is coupled only to one of the two-qubit levels, such that the Rydberg pulses have no effect for qubits in the other state. Combined with the Rydberg blockade that prevents simultaneous excitation of two atoms within a certain spatial range, the logic of a \textit{CZ}-gate can be mapped. 

The spatial range of the Rydberg blockade, the \emph{blockade radius}, defines the connectivity of the qubit register: Each qubit can perform a two-qubit gate with any other qubit lying within its blockade radius. Typical values for the blockade radius are~$2$ to~$3$ lattice sites, which corresponds to a next-nearest-neighbour connectivity in a square lattice.  In principle, however, also higher connectivities up to~1:50 are possible. The size of the blockade radius, among other things, depends on the energy of the Rydberg state and the intensity of the exciting lasers. On the other hand, higher-lying Rydberg states are more susceptible to noise, because of their strong dipole moment.

Using the mobile optical traps, it is also possible to rearrange neutral atom qubits during the computation process to enable dynamic, non-local entanglement. In this approach, which is described in~\cite{bluvstein_quantum_2022}, qubits are arranged in pairs with small relative distances. Each of these pairs gets entangled by a global \textit{CZ}-gate directed to all atoms within a certain area. Then, the pairs are rearranged using the mobile traps and another global \textit{CZ}-gate is performed, such that previously entangled atoms are now additionally entangled with other atoms. In order to keep the coherence of the qubit state, the movement can only be performed with a limited speed, which, however, is still sufficient to transport information over a spatial range of about~$2000$ qubits.  

The qubit scheme described here, which encodes the states~$\ket{0}$ and~$\ket{1}$ in the ground state manifold and uses the Rydberg excitation for two-qubit gates, is called \emph{gg}-scheme. It is also possible to encode the qubit states into one ground state and a Rydberg state (\emph{gr}) or in two different Rydberg levels (\emph{rr})~\cite{Morgado_Whitlock_2021}. While the latter two approaches imply shorter coherence times, gate operations can often be realised faster.

\subsection{Decoherence and noise}\label{sec:decoherence}

There are various different sources of noise which can disturb cold-atom qubits and degrade their coherence. Most prominently, the laser sources used for cooling, trapping, addressing and readout can suffer from laser intensity or phase noise. Moreover, as mentioned above, residual atoms in the vacuum cell limit the qubit lifetime due to atomic collisions. Another source of noise are (time-dependent) electric and magnetic fields arising from devices in the laboratory or external sources such as the earth's magnetic field.  

The robustness of a qubit against noise is usually expressed by longitudinal~($T_1$) and transversal~($T_2$) relaxation times. The~$T_1$ time measures the decay time from the~$\ket{1}$ state to the~$\ket{0}$ state.
This measure captures the effect of dissipation and provides only insight into the stability of an eigenstate of a qubit. 
The stability of the relative phase of a superposition state is expressed by the so-called \emph{dephasing times}.
The~$T_2$ time is the time after which, with probability~$1/e$, an initial state~$\ket{+}=\left(\ket{0} + \ket{1}\right)/\sqrt{2}$ evolves into an equal mixture of the~$\ket{+}$ and~$\ket{-}=\left(\ket{0} - \ket{1}\right)/\sqrt{2}$ states. In the picture of the Bloch sphere this means that the projection of the vector onto the axes perpendicular to~$\ket{0}$ and~$\ket{1}$ shrinks. For an ensemble of multiple qubits, due to small variations in the energy difference between the~$\ket0$ and~$\ket1$ states, the relative superposition phases can evolve differently for several qubits. This dephasing effect is measured by the inhomogeneous transversal relaxation time~$T_2^*$.

\subsection{Scalability and technical challenges}\label{sec:challenges}

Currently available neutral atom quantum computers and simulators have in the order of~$100$ qubits. One of the main challenges in the field of neutral atom quantum devices is to scale up the systems to larger number of qubits while at the same time improving their coherence. The number of traps in the described optical trap arrays is mainly limited by the available laser power and the performance of the optical system that is used to generate them~\cite{henriet_quantum_2020}, promising good scalability of up to several thousands of qubits.

However, when increasing the number of qubits, the necessary rearrangement of the atoms prior to the computation becomes more complex and will take more time. In order not to loose an atom during the transport process, the speed of transport is limited. On the other hand, all moves have to be done within the lifetime of an atom inside the trap. Moreover, the transfer probability between the static and moving traps is reduced by collisions with residual gas and inaccuracy in positioning the tweezers~\cite{schymik_enhanced_2020}. In general, the rearrangement process needs to be engineered such that the number of moves and the distances to travel are minimised.  

The lifetimes inside the traps can be increased by further reduction of the residual pressure inside the vacuum cell, which is facilitated by lowering the overall temperature: when cooling the experimental setup to about~$4\,\unit{K}$, trap lifetimes of up to~$6000\,\unit{s}$ have been observed in experiments~\cite{schymik_lifetime_2021}. The installation of a cryostat, however, leads to more infrastructure constraints and makes the miniaturisation of the setup more difficult. 

The depth of the traps can be increased by changing the frequency detuning of the laser light to be closer to the atomic transition. As this in turn leads to shorter $T_1$~decoherence times, a trade-off between these effects has to be found. 

Another important challenge is the finite lifetime of an atom in the Rydberg state, which can give rise to different kinds of correlated errors that are difficult to address by standard error correction methods~\cite{cong_hardware-efficient_2022}. The Rydberg lifetimes lie typically in the order of~$150~\unit{\micro s}$~\cite{bluvstein_quantum_2022} and are mainly limited by spontaneous radiative decay and transitions induced by black-body radiation. The latter can also be greatly reduced by cooling of the setup, since the black-body radiation intensity scales as~$T^4$~\cite{schymik_lifetime_2021}.

Apart from optimising the preparation and rearrangement, the overall speed of the computation process could also be improved by executing more gates in parallel. As described above, in most of the current setups only the same single qubit gate can be executed on several atoms at the same time. In future setups, a higher degree of parallelisation could be achieved by the installation of several addressing beams with different parameters. Moreover, there are efforts to speed up the gate operations themselves, such as the ultrafast interaction observed for a single qubit in~\cite{chew_ultrafast_2022}.

The fidelities of the two-qubit gate operations also need to be further improved, the highest demonstrated fidelity is~$\sim 0.995$~\cite{evered_high-fidelity_2023}. Some of the main challenges are finite atomic temperatures and off-resonant laser scattering. These could be met by further laser cooling of the atoms inside the traps or by using higher laser powers~\cite{levine_parallel_2019}. Another approach to reach the desired quantum state with higher probability is to optimise the pulse shapes in time instead of using standard square-shaped pulses~\cite{henriet_quantum_2020}. The readout-fidelities can also be improved, for instance, by using so-called non-destructive readout protocols, as mentioned in~\cite{levine_parallel_2019}.

\section{Overview of the neutral atom platform}

After presenting how a cold-atom quantum computer works, we now focus on a set of properties that describe its computing capabilities. We start with a discussion of the properties on a level that is very close to the hardware itself. Afterwards we extend our viewpoint to a more application-oriented perspective.

\subsection{Relevant hardware properties}\label{subsec:explanation_criteria}

Defining the right metrics to properly represent the performance of diverse types of quantum systems is critical to both users and developers of a quantum computing system. In this paper, a set of parameters has been carefully selected that is general enough to characterise different quantum computing platforms while maintaining enough detail to assist decisions on the platform. The parameters are described in more detail below and typical values for the neutral atom platform are given in Tab.~\ref{tab:Parameters_1}. 

\subsubsection{Qubits}

The most obvious criterion to characterise any quantum computing platform is the amount of available qubits and the possibility to increase it further in the future. 
The qubit connectivity describes the number of qubits that one qubit can directly interact with. Thus, it is a measure for the capability to perform entangling operations. Missing interactions between qubits can be accounted for by inserting \textit{SWAP}-gates. For neutral atom quantum computers, the connectivity depends mainly on the energy of the Rydberg state, the intensity of the exciting lasers, and the location of the qubit in the array (see Sec.~\ref{sec:tq-gates}). 

Another interesting aspect is whether a certain hardware platform can realise systems with more than two states, so-called \emph{qudits}. These have~$d$ different states (so,~$\ket{0}, \ket{1}, \ket{2},..,\ket{d-1}$) instead of only~$2$ as for qubits. By increasing the number of states, more information can be encoded in the same number of particles. Since the ground and excited states of neutral atoms naturally exhibit more than two energy levels, the realisation of qudits is possible and is currently being explored, also in the context of quantum memories~\cite{dong_storage_2023}. 

\subsubsection{Lifetimes and decoherence times}

The lifetimes and decoherence times determine, amongst other parameters, how many operations can be performed on a qubit. In the context of cold atoms, the lifetime denotes the average lifetime of an atom in an optical trap, either an optical tweezer or an optical lattice, as described in Sec.~\ref{sec:trapping}. The decoherence times give information about how long the state of a qubit is conserved until it decays. The exact definitions of the~$T_1$,~$T_2$ and~$T_2^*$ times are given in Sec.~\ref{sec:decoherence}. In general, the trap lifetimes, as well as the decoherence times, differ with respect to the specific atomic energy levels which are chosen as qubit states: As mentioned in Sec.~\ref{sec:qubit_encoding}, encoding the qubit in the nuclear spin levels, for instance when using Strontium atoms, increases the decoherence and lifetimes, which is also reflected by the values in Tab.~\ref{tab:Parameters_1}.

\subsubsection{Native gates}

Every quantum computer needs to be able to realise a universal gate set, from which all operations can be generated. However, the efficiency of a certain quantum computing platform can be greatly increased, if more gates can be implemented natively, such as \textit{SWAP}-gates or gates involving more than two qubits. This reduces the overall gate overhead when compiling a quantum program to the specific hardware device being used. 
Moreover, the parallel execution of gates can speed up the calculation on a quantum computer significantly. As mentioned in Sec.~\ref{sec:gates}, this is possible to a certain extent for neutral atoms by using several addressing beams or global Rydberg pulses which entangle all atoms being close enough to each other.

\subsubsection{Fidelities of operations}

The quality of a gate operation is usually characterised by its fidelity, which is defined as the overlap of the qubit state after the gate execution with the ideal qubit state after the desired operation. If the fidelity of a certain gate operation is determined, usually, the value is corrected afterwards to account for imperfections in state preparation and measurement (SPAM). The measurement or read-out of the qubit states is often also characterised by a fidelity, which quantifies how often, for instance, the value~$0$ is measured if the qubit was in the state~$\ket{0}$. The preparation of the qubits before the start of the computing sequence also influences the performance, since the quantum register needs to be initialised in a well-defined state. As described in Sec.~\ref{sec:quantum_computing_sequence}, rearrangement of the qubits into a homogeneously filled array is needed for neutral atom quantum computers prior to computation. The initialisation efficiency can be quantified by the success probability of the rearrangement process. 

\subsubsection{Execution times}

The execution time of quantum gate operations on the one hand determines the number of operations that can be performed within the decoherence time and on the other hand influences the overall speed of a quantum computation.  To specify the speed of the neutral atom platform in particular, we list the execution times for the bare quantum operations such as gates, measurement and preparation in Tab.~\ref{tab:Parameters_1}.

\subsubsection{Installation and operation}

Most quantum computers will be used in a hybrid setup alongside classical computers. Thus, the infrastructure requirements for a certain quantum computing platform need to be considered in view of integrating it, for instance, with a classical supercomputer or installing it at the user's facility such as a production shop floor.
Moreover, depending on the type of quantum computer, calibrations and measurements of qubit properties might be necessary prior to operation. This prolongs the overall execution time.

In view of running more complex algorithms that involve many qubits and, for instance, storing of quantum information during runtime, variable qubit registers will be convenient. A special property of the neutral atom platform is the ability to shuttle atoms during the computation (see Sec.~\ref{sec:tq-gates}), which enables entanglement between previously distant qubits and allows for the definition of specific zones, for instance, for computation or storage of information. 

As a last point, we list the different paradigms or models of quantum computation that can be realised with the neutral atoms platform. So far, we have mainly discussed properties important for gate-based or digital quantum computation. However, using neutral atoms, also a combination of gates and analogue blocks (digital-analogue quantum computing), quantum annealing or analogue, as well as digital quantum simulations, are possible.

\begin{table*}[htbp]
    \centering  
    \begin{tabularx}{\linewidth}{p{5cm}X}
       \bf{Parameter}  &\bf{Typical values today (near Future)}\\
       \hline
       \hline
       \noalign{\vskip 1mm}    
       \multicolumn{2}{l}{\bf{Qubits}}\\
        Amount & $\sim 100$~\cite{henriet_quantum_2020, bluvstein_quantum_2022, graham_multi-qubit_2022, barnes_assembly_2022} ($\sim 1000$ for 2024~\cite{pasqal_roadmap_2023, quera_tech_review_2021})  \\
        Connectivity & $10:1-20:1$~\cite{henriet_quantum_2020}, ($50:1-100:1$ possible in principle) \\
        Multiple states (i.e., qudit) &  In principle possible  \\
        \noalign{\vskip 0.5mm} 
        \hline
        \noalign{\vskip 1mm} 
        \multicolumn{2}{l}{\bf{Lifetimes and Decoherence times}}\\
        Trap lifetime & $10-60\,\unit{s}$~\cite{ebadi_quantum_2021, graham_multi-qubit_2022, schymik_enhanced_2020, barnes_assembly_2022} (up to~$6000\,\unit{s}$ with cryostat~\cite{schymik_lifetime_2021})\\
				Decoherence times (electronic spin) & $T_1 \sim 4\,\unit{s}$,~$T_2 \sim 1\,\unit{s}$,~$T_2^* \sim 4\,\unit{ms}$~\cite{bluvstein_quantum_2022, graham_multi-qubit_2022} \\
				Decoherence times (nuclear spin) & $T_1 \gg 5\,\unit{s}$,~$T_2 \sim 40\,\unit{s}$,~$T_2^* \gg 3\,\unit{s}$~\cite{barnes_assembly_2022}\\
        \noalign{\vskip 0.5mm} 
        \hline
        \noalign{\vskip 1mm} 
        \multicolumn{2}{l}{\bf{Native gates (\textit{gg}-qubits)}}\\
				List of gates & Single-qubit rotations, \textit{CZ}~$\rightarrow$ \textit{CNOT}~\cite{henriet_quantum_2020, levine_parallel_2019}, \textit{SWAP, CPHASE}~\cite{Morgado_Whitlock_2021} \\
       $>$2-qubit gates: & \textit{CCZ}~$\rightarrow$ \text{Toffoli /}\textit{CCNOT}~\cite{henriet_quantum_2020, levine_parallel_2019},~$\textit{C}_k\textit{Z}$, generalisation to~$k$ control qubits~\cite{Morgado_Whitlock_2021} \\
       Parallelism & Apply the same gate on multiple Qubits simultaneously: Single-qubit rotations, \textit{CZ}~\cite{bluvstein_quantum_2022} (Multiple gates on multiple Qubits) \\
        \noalign{\vskip 0.5mm} 
        \hline
        \noalign{\vskip 1mm} 
        \multicolumn{2}{l}{\bf{Fidelities of operations}}\\
        1-qubit gate  & $0.996-0.999$~\cite{bluvstein_quantum_2022, graham_multi-qubit_2022}  \\
        2-qubit gate  & $0.955 - 0.995$~\cite{levine_parallel_2019, graham_multi-qubit_2022, evered_high-fidelity_2023} \\ 
        Readout & $\gtrsim 0.95$~\cite{xu_preparation_2021}	 \\
				Preparation & Trap occupation probability (after rearrangement):~$0.988$~\cite{endres_atom_2016} \\ 
				& Success probability for defect-free array~$\sim 0.75$, depending on size of array~\cite{endres_atom_2016} \\	
        \noalign{\vskip 0.5mm} 
        \hline
        \noalign{\vskip 1mm} 
        \multicolumn{2}{l}{\bf{Execution times}}\\
        1-qubit gate & $\sim 2\,\mu \unit{s}$ ($\pi$-pulse)~\cite{bluvstein_quantum_2022} \\ 
        2-qubit gate & $\sim 400\,\unit{ns}$ (\textit{CZ})~\cite{bluvstein_quantum_2022, levine_parallel_2019} \\
        Preparation (incl. rearrangement) & $\sim 400\,\unit{ms}$~\cite{endres_atom_2016} \\ 
        Readout & $\sim 10\,\unit{ms}$ for fluorescence imaging~\cite{xu_preparation_2021}, (~$\sim 6\,\mu \unit{s}$ using a collective readout scheme)~\cite{xu_preparation_2021}  \\
        \noalign{\vskip 0.5mm} 
        \hline
        \noalign{\vskip 1mm} 
        \multicolumn{2}{l}{\bf{Installation and operation}}\\
        Required infrastructure & Vacuum cell and pumps, lasers, optical elements, microwave sources, signal generators and modulators, magnetic field coils. Cooling the setup with a cryostat can improve vacuum quality and increase trap lifetimes~\cite{schymik_lifetime_2021}. (Rack-level implementation possible~\cite{pasqal_hardware_2023}) \\
        Calibration & Rearrangement at beginning, no calibration of individual qubits necessary  \\
        Specificity & Shuttling operations~\cite{bluvstein_quantum_2022} \\
        Access & Via the cloud~\cite{azure_quantum_2023, aws_braket_2023} and on-premise  \\
				Quantum computing paradigm &  Gate-based (digital) quantum computing, digital-analogue quantum computing, quantum annealing, analogue quantum simulation \\
        \hline
        \hline
    \end{tabularx}
    \vspace{0.1cm}
    \caption{Summary of parameters for neutral atom quantum computers, representing the current state-of-the-art with an extrapolation to the future given in brackets. The values reported here reflect the status of the devices run by neutral atom quantum computing companies, with the perspective of being used for applications already now or in the near future. A description of each parameter can be found in Sec.~\ref{subsec:explanation_criteria}.} 
    \label{tab:Parameters_1}
\end{table*}

\subsection{Significance for applications}\label{subsec:sig_applications}
The various properties of neutral atom quantum computing hardware described and listed above directly influence the performance of quantum algorithms being executed on them. Especially in the current era of NISQ-devices, the details of the hardware, but also of the problem to be solved are essential when developing software. This is true for analogue quantum simulators as well as for gate-based approaches, and applies to data loading, system preparation as well as manipulation and readout. In the following, we discuss which of the parameters listed above are most relevant for running applications and for implementing quantum error correction schemes.   
    
\subsubsection{Influence of hardware properties on the performance of quantum algorithms}

For NISQ-devices, low-level properties like the measures listed in Table~\ref{tab:Parameters_1} can give an indication whether a certain quantum computing hardware platform is suitable to solve a given task. For gate-based quantum computing, which is the most common scheme implemented with cold atoms, the number of qubits is directly related to the size of the problems that could be tackled. 

The number of gates needed to implement a certain quantum algorithm on a specific quantum processing unit (QPU) depends on the connectivity between the qubits and the available native gate set. Operations like a \textit{SWAP}-gate, which swaps the states of two qubits and appears when compensating for limited connectivity, can be costly, if they need to be further decomposed into a native gate set. Thus, when solving problems that require high connectivity, the circuit depth increases substantially due to the insertion and decomposition of \textit{SWAP}-gates. In general, the required qubit connectivity of a quantum hardware device can be determined by the number of connections in the graph that represents the executions of two-qubit gates in the problem-specific quantum circuit. Thus, it is not always necessary to have an all-to-all connectivity~\cite{wintersperger:22:codes} and choosing a platform with lower connectivity but other advantages such as a larger number of qubits would be optimal.

Apart from their relatively high connectivity, neutral atom quantum computers feature the native implementation of \textit{SWAP}-gates, which reduces the overhead in gate count and circuit depth. Since every gate operation introduces an error and prolongs the overall execution time, low gate counts are significant in the NISQ-era. Moreover, the available native implementation of gates with more than two qubits can help to run certain quantum algorithms more efficiently, such as Grover's search algorithm or solving non-linear partial differential equations~\cite{henriet_quantum_2020}.

Apart from reducing the overhead in \textit{SWAP}-gates, higher connectivity between the qubits is also desired for implementing efficient error-correction schemes such as low-density parity check (LDPC) codes which in turn require lower numbers of physical qubits~\cite{bravyi_future_2022}. 

The fidelities of the different gates directly determine the maximum number of gates that can be executed on a certain quantum device. But also in view of implementing error correction, the fidelities of the gate and also measurement operations are important, as the detection and subsequent corrections of errors relies on these operations. The measurement errors can also influence the performance of hybrid quantum algorithms which involve repeated measurements during the runtime of the program. 

The execution time of a quantum algorithm also depends on the properties of the hardware, most importantly on the duration of the preparation, gate and measurement operations. The overall runtime of a quantum computation is often characterised by the so-called \emph{wall-clock time}, that is, the end-to-end execution time from a user perspective, including data load, compilation of quantum circuits to the specific hardware, execution and readout and communication times. The wall-clock time does not only depend on the type of QPU, but also on the connection between the QPU and the user and on the properties of classical computers involved in the pre- and post-processing of data, which are not specific to a certain type of quantum hardware.

\subsubsection{Fault-tolerance and error correction}

To utilize quantum computing for large-scale problems it needs to be fault-tolerant and error-corrected. As quantum computers are inherently sensitive to noise, errors are likely to occur and due to interactions between the qubits they will propagate through the circuit, which can lead to a cascade of errors. Fault-tolerant quantum computing describes a set of methods to correct errors introduced by unwanted interaction with the environment or faulty control (Quantum Error Correction, QEC) and to design quantum circuits such that QEC can be implemented and errors do not cascade through the circuit. Assuring the latter, a threshold can be achieved at which QEC is correcting more errors than are being created and the computation can be scaled up, provided that the fidelites of individual operations and qubits are high enough~\cite{Paler_2015}.  

%Quantum error correction (QEC) refers to a set of methods to detect and correct errors that can occur in a quantum computer. 
In contrast to classical bits, qubits cannot only suffer from bit-flip errors, but also phase errors can occur, which means that the relative phase of a superposition state is changed. Even a slight change of the phase can lead to a wrong result. This analogue nature and the no-cloning theorem make QEC more complex than its classical counterpart, which relies on copying the information stored in a bit~\cite{Gottesman_2010}. 
In QEC, the information of one logical qubit is encoded in the states of several physical qubits. Depending on the scheme being used, typically between~$7$~\cite{Steane_1996} and~$9$~\cite{Bacon_2006, Bombin_2007, Horsman_2012} physical qubits are needed to represent the state of one qubit. To detect errors and perform corrections, often additional auxiliary qubits are being used. Thus, the actual number of physical qubits required for QEC is increased further~\cite{Chamberland_2018, chao_2020}. Moreover, if several layers of encoding are implemented, the qubit overhead multiplies and up to thousands of physical qubits might be needed to encode one logical qubit~\cite{chao_fault-tolerant_2018}. 

%This representation allows to take measurements on some of the physical qubits to detect errors without directly measuring, and thus projecting the encoded state. The result of these syndrome measurements can be stored using additional auxiliary qubits, and depending on their state the appropriate error correction operation can be performed. 
%To realise error correction in practise, the protocol needs to be \emph{fault-tolerant}, which means that the syndrome measurements and correction operations do not disturb the protection of the code, even when considering that the gate operations themselves are noisy. The code should still produce the correct results, even when single components of the circuit fail. This goal can be achieved, provided that the error rates of the gate operations lie below certain threshold values~\cite{Gottesman_2010}. 
While general error correction codes have the goal to protect against all possible errors, not all types of errors occur in every quantum device or at least not with the same probability. Thus, the number of qubits needed for error correction might be greatly reduced by using schemes adapted to a specific hardware platform, taking into account its typical sources of errors. For cold atoms, errors due to the decay of Rydberg states into levels outside the computational subspace are an important challenge that is not covered by generic error correction methods. Those have been addressed recently by a hardware-efficient, fault-tolerant error correction scheme, making use of the specific properties of Rydberg atoms~\cite{Iris2022}. 
Thus, knowledge about the error sources and mechanisms of the used quantum computing hardware platform is essential for the design and implementation of efficient fault-tolerant codes. 

Another important issue in this context is cross-talk, which describes any unwanted interaction between qubits or between qubits and the control signals: a gate pulse
can affect other than the intended target qubit(s) or local gate operations are disturbed by other gate operations applied in parallel. Cross-talk can break fault-tolerant circuit constructions that rely on the assumption that errors occur only on those qubits which are being addressed by gate operations. In neutral atom quantum computers, the amount of cross-talk is usually moderate~\cite{Xia_2015, Graham_2019}, since the distances between single atoms are usually large enough compared to the focus size of the addressing lasers, avoiding unwanted excitation. 

\subsection{Overview of neutral atom providers}~\label{subsec:providers}

Multiple companies and start-ups building and researching the neutral atom quantum computing platform exist. Some already provide access to available hardware either via in-house managed access systems or third-party cloud providers. The company Pasqal offers a digital-analog quantum processor with~$100$ qubits, based on optical tweezer arrays and Rydberg interactions~\cite{pasqal_2023}, which is available on Azure Quantum ~\cite{azure_quantum_2023}. Another quantum processing unit being accessible via common cloud providers, in this case, AWS Braket~\cite{aws_braket_2023}, is the Aquila device by QuEra. It operates up to 256 qubits in analogue mode, also utilising Rydberg interactions and optical tweezers with a variable geometric layout~\cite{quera_2023}.
ColdQuanta offers a different type of analogue quantum simulator based on a Bose-Einstein-condensate, that is accessible via their own cloud platform~\cite{coldquanta_albert_2023}. A gate-based quantum computer is planned to be made available in the near future~\cite{coldquanta_hilbert_2023}.
Other vendors such as AtomComputing~\cite{atomcomputing_2023} and Planqc~\cite{planqc_2023} are currently planning and building first prototypes of quantum computers, which also might be available in the next years, most probably also via cloud access.

\section{Discussion}\label{sec:discussion}

In the previous sections we presented important properties of QPUs and their relation to performance of algorithms and implementation of error correction. As any quantum computing hardware platform today, also neutral atoms have certain advantages as well as disadvantages, which will be discussed in this section, also comparing with other mature platforms such as superconducting qubits and ion traps. In Sec.~\ref{sec:algorithms}, we describe applications for which neutral atom quantum computers have already been employed successfully or which might be well-suited for this platform. Finally, a conclusion of our review is provided. 

\subsection{Summary and comparison with other platforms}

One of the most important advantages of neutral atoms is the good scalability compared to other platforms such as ion traps but also superconducting qubits. Devices with 100 and more qubits are available already today and these can be scaled up to multiple thousands of qubits using a single trap array or lattice. Alike other platforms, also 'horizontal scaling', that is, connecting multiple QPUs is possible. Moreover, one has to view the properties of the platform relative to its maturity: while analogue simulation with ultra-cold atoms in optical lattices has been researched for decades now, the field of digital quantum computing with neutral atoms is younger, especially compared with ion traps and superconducting qubits.
Another interesting feature is the intermediate connectivity between qubits, which can be made variable by shuttling atoms during the computation, similar as for ion traps, which, however, feature all-to-all connectivity.
For superconducting qubits, schemes to extend the connectivity exist and have been tested, but issues such as higher cross-talk have to be considered~\cite{stassi_scalable_2020, bravyi_future_2022}, whereas cross-talk is less of a problem in neutral atoms.

The duration of the gate operations for neutral atoms are in a similar range as for ion traps, but slower than superconducting qubits. However, this is compensated by the relatively long coherence times, especially for approaches using nuclear spin qubits such as in Strontium or Ytterbium~\cite{Jenkins_2022, Ma_2022}. Moreover, similar to ion traps, the qubits are identical by nature, so no calibration of individual qubits is necessary and the influence of noise is homogeneous across all qubits.    

While the error-rates for single-qubit gates are already quite low, the highest two-qubit gate fidelity measured is $\sim 99.5\%$, as shown in Tab.~\ref{tab:Parameters_1}. However, this value already surpasses threshold for quantum error correction~\cite{Fowler_2012} and lies in a similar range as for other platforms such as superconducting qubits, while ion traps currently feature higher two-qubit gate fidelities.
In view of implementing fault-tolerant codes, higher gate fidelities also lead to a reduction of the overhead of physical qubits needed for encoding~\cite{Suchara_2013}. 
As mentioned above, the field is still comparably young and the further improvement of the gate fidelities is a prime target for most neutral atom vendors. There are no fundamental reasons why the gate fidelities should not reach higher values, since they are mostly limited by technical issues such as finite atomic temperatures and off-resonant laser scattering, as described in Sec.~\ref{sec:challenges}. 

While the gate operation times are acceptable compared to the decoherence times, the long preparation times prior to computation are another unfavourable factor. The rearrangement process will become more complex when scaling to higher numbers of qubits and thus might take even more time. Moreover, as opposed to superconducting qubits and other solid state systems, neutral atom qubits are lost after each measurement and the complete preparation process including cooling and rearrangement needs to be repeated over again. Nevertheless, there are efforts to improve on these issues: schemes for near-deterministic loading of optical tweezers have been realised in research labs~\cite{Brown_2019, Jenkins_2022, Ma_2022}, to overcome the stochastic nature of the trap loading and to finally supersede the rearrangement process. 

As mentioned in Table~\ref{tab:Parameters_1}, the miniaturisation of cold atom setups to the level of 19-inch racks is possible and is already being implemented~\cite{pasqal_hardware_2023}. The required infrastructure such as lasers, vacuum pumps or microwave generators is not particularly complex and cold atom experiments have already been implemented under challenging environmental conditions~\cite{aveline_observation_2020}. Using integrated optics, as in trapped ion setups, can also help to make the setups smaller and more robust. 
Although cryostats might be used in the future to increase the qubit lifetimes, providing on-premise access to neutral atom quantum computers or integrating them into classical supercomputing centres seems to be well possible in the near future.

\subsection{Applications of neutral atom quantum computers}~\label{sec:algorithms}

The big questions for identifying potential added value of having a QPU or a quantum computing component in the end-to-end process for problem solving definitely depend on the problem at hand. Is it possible to formulate the problem in a way that it would benefit from quantum components, which use specific quantum features in their calculation process? Does the orchestration and overall architecture of the quantum computing platform suit to the problem class? In the following, we describe several examples of problems that have already been tackled successfully by neutral atom quantum processors or are proposed to be good candidates.

\subsubsection{Quantum simulation for material science and physics problems}

Analogue quantum simulation with ultra-cold neutral atoms has been performed in research labs since many years and is being used for various applications ranging from statistical physics~\cite{greiner_quantum_2002} and material science, such as the investigation of high-temperature superconductors~\cite{Cheuk_2016, Boll_2016, Parsons_2016, Brown_2017, Hague_2020, hirthe_magnetically_2023}, or simulation of spin systems~\cite{eisert_quantum_2015, Lienhard_2018, Semeghini2021, Keesling2018, Bernien2017} to high-energy physics and astrophysics~\cite{Giovanazzi_2005, Zohar_2016, nachman_quantum_2021}.
In analogue quantum simulation, the problem is being simulated by tailoring the Hamiltonian of the qubit system to directly mimic the Hamiltonian of the problem and study its behaviour under the influence of this Hamiltonian. For simulations in material science, the atoms typically mimic the electrons in the material and an optical lattice is being used to represent the crystal lattice in which the electrons can move. Today, an extended toolbox for this type of simulations has been developed~\cite{schafer_tools_2020}, including also periodic driving of the system to realise topological phases of matter~\cite{Oka_2019}. Nevertheless, the range of applications that can be simulated in this way remains limited to those which match the symmetry and capabilities of the qubit system~\cite{henriet_quantum_2020}, so analogue quantum simulation is not universal. Using digital quantum simulation based on gate operations, all types of problems can be tackled in principle. 

\subsubsection{Combinatorial optimisation problems}

Quantum computers based on neutral atoms and Rydberg interactions are seen as promising candidates for solving combinatorial optimisation problems. A specific class of problems that has been implemented in a hardware-efficient way is the Maximum independent set (MIS) problem on unit disc graphs, since the independent set constraint is natively fulfilled by the Rydberg blockade mechanism~\cite{ebadi_MIS_2022}. There are several algorithmic classes which can be used to solve the MIS on a neutral atom platform, such as the Quantum Approximate Optimisation Algorithm (QAOA), quantum annealing algorithms (QAA)~\cite{Pichler_MIS_2018}, or variational quantum simulation as described in~\cite{ebadi_MIS_2022, Serret_2020}. In the latter approach, the parameters of the Hamiltonian describing the Rydberg drive and interaction are optimised by a classical optimiser such that the state after the evolution minimises the expectation value of the problem Hamiltonian. In this way, the ground state encoding the optimal solution is approximated by (a sequence of) global Rydberg pulses applied to the atoms, that in turn are arranged in a pattern that matches the problem graph in the case of MIS. 
The MIS problem on a unit disk graph has various applications in finance, bioinformatics, logistics, and automation~\cite{Wurtz_2022}.
Moreover, the scheme to solve this subclass of MIS can in principle be extended to maximum weighted independent set problems on unit-disk graphs which include maximum weighted independent set on graphs with arbitrary connectivity, quadratic unconstrained binary optimisation problems with arbitrary or restricted connectivity, and integer factorisation~\cite{Nguyen_2023}. A proposal to extend this scheme to general optimisation problems, also containing higher-order terms, is described in~\cite{Lanthaler_2023}. 

In view of tackling general combinatorial optimisation problems, schemes to implement quantum annealing with neutral atoms have been proposed. While some parts of the standard annealing algorithm like the initial state~$\ket{+}^{\otimes n}$ or the hard-sphere interaction in the Ising model are difficult to implement with Rydberg atoms~\cite{Serret_2020}, several approaches exist to overcome this challenges such as the stroboscopic scheme proposed in~\cite{Pichler_MIS_2018} or optimising the locations of the atoms which also influences the strength of the interaction. A general proposal for quantum annealing with neutral atom processors utilises the so-called \emph{LHZ}-scheme~\cite{Lechner_2015, glaetzle_coherent_2017}, in which the qubits represent the edges of a problem graph, allowing to realise an effective all-to-all connectivity with local Rydberg interactions. This approach can also be used to overcome the limited connectivity of neutral atoms QPUs in the context of gate-based algorithms such as QAOA, enabled e.g., by the use of four-body Rydberg gates~\cite{Dlaska_2022}.

\subsubsection{Differential equations and machine learning}
Neutral atom quantum computers may also be suitable to solving other types of
industry-relevant problems such as solving differential equations~\cite{henriet_quantum_2020}.
One example is so-called \emph{Physics-informed machine learning} (PIML),
a class of universal function approximators that is capable of encoding any
underlying physical laws that govern a given data-set, and can be described
by partial differential equations~\cite{Raissi_Perdikaris_Karniadakis_2019}.

Machine learning is among the most prolific fields of research in the
quantum computing community, and many machine learning techniques (see, \eg, 
Refs.~\cite{Ciliberto2018,abo20Class,Zhao21QNN,Biamonte2017,Meyer:2022}) have 
corresponding equivalents in the quantum domain. Exponential speedup has, for
instance, shown to be possible for quantum support vector 
machines~\cite{Rebentrost:2014}, while quantum classifiers~\cite{Havlicek:2019}, which 
have even been implemented on small-scale NISQ machines, can derive (classically) hard to estimate kernels. Regardless of the conditions of their advantages, these approaches need careful consideration of subtle issues (see, \eg, 
Refs.~\cite{Tang:2021,Skolik:2021,Franz:22,McClean:2018}) 
that do not appear in classical algorithms and are, in particular, strongly
intertwined with the underlying hardware. As for concrete \emph{practical}
applications, suffice it to say that these are still unknown to the best of our
knowledge.

Pasqal \href{https://medium.com/pasqal-io/neutral-atom-quantum-computing-for-physics-informed-machine-learning-1f5ee4c055a6}{suggests} 
that variational techniques ubiquitous in quantum machine learning (QML) may be extended to
physics-informed machine
learning~\cite{Raissi_Perdikaris_Karniadakis_2019}, which 
could be beneficially implemented on cold atom systems. Yet, and 
with the other notable exception of Ref.~\cite{Gianani:2022}, the available
body of literature on QML on cold atom machines is astonishingly scarce, 
to the best of our knowledge (interestingly, classical machine learning has
been used to improve experiments with ultra-cold atoms~\cite{Wigley:2016}).

Regarding software frameworks, commercial access to cold atom hardware is still more cumbersome than for other hardware platforms
of large commercial providers. Even if the relevance of \emph{which} software 
abstraction layer is used is of subordinate importance~\cite{schoenberger:22:icsa},
the general support for cold atom quantum computers in QML frameworks like 
PennyLane~\cite{Bergholm:2022} or Tensorflow  Quantum~\cite{Broughton:2021}, which are 
often used in exploratory research, does not seem to be en par with hardware offered by 
large commercial vendors. For instance, Cirq support, which Tensorflow quantum is based 
on, has been announced for Pasqal hardware \cite{Google:Pascal},
yet is not available to the general public at the moment. Likewise, Pennylane does not support cold atom backends
at the time of
writing \cite{PennyLane:Plugin}, while efforts to support cold atom architectures are,
for instance, available as an open source plugins
\cite{PennyLane:git}.

\subsection{Conclusion}

In this work we provided a detailed insight into the strengths and weaknesses of neutral atom quantum computers and illustrated how their low-level properties influence the performance of quantum algorithms. In the current NISQ-era, the suitability of quantum computers for a specific problem cannot be captured by general benchmark metrics alone, but the details of the hardware must be taken into account and algorithms need to be designed in a problem-specific and hardware-efficient way. 
While cold atoms have been used for analogue quantum simulation for a longer time, performing digital quantum computing is a comparably young field of research. There are still some drawbacks such as the long preparation times and other technical challenges. Nevertheless, the neutral atom platform has promising advantages such as good scalability, connectivity, and tunability regarding the qubit arrangement and computing mode and is expected to catch up with other established platforms in the upcoming years. 

\begin{backmatter}

\section*{List of abbreviations}
\begin{table}[h!]
      \begin{tabular}{c|c}
        \hline
        NV &  Nitrogen-vacancy \\ \hline
         NISQ  & Noisy intermediate-scale quantum \\ \hline
         HPC & High-performance computing \\ \hline
         CLOPS & Circuit layer operations per second \\ \hline
         KPI & Key performance indicator \\ \hline
         MOT & Magneto-optical trap \\ \hline
         SLM & Spatial light modulator \\ \hline
         AOD & Accousto-optical deflector \\ \hline
         CCD & Charge-coupled device \\ \hline
         SPAM & state preparation and measurement \\ \hline
         QPU & quantum processing unit \\ \hline
         LDPC & low-density parity check \\ \hline
         QEC & Quantum error correction \\ \hline
         MIS & Maximum independent set \\ \hline
         QAOA & Quantum Approximate Optimisation Algorithm \\ \hline
         QAA & Quantum annealing algorithm \\ \hline
         PIML & Physics-informed machine learning \\ \hline
         QML & Quantum machine learning\\ \hline
      \end{tabular}
\end{table}

\section*{Ethical Approval and Consent to participate}
Not applicable.

\section*{Consent for publication}
Not applicable.

\section*{Availability of supporting data}
Not applicable.

\section*{Competing interests}
The authors declare that they have no competing interests.

\section*{Funding}
Not applicable.

\section*{Author's contributions}
KW, TE, TS, SL, FD, WM and JK prepared the initial manuscript, KW provided details on the neutral atom platform. All authors read and approved the final manuscript. All work was conducted in the context of the Quantum Technology and Application Consortium (QUTAC).

\section*{Acknowledgements}
  We would like to thank the following cold-atom quantum computing providers listed in alphabetical order for the time they spent with us discussing their quantum computing platform: AtomComputing, ColdQuanta, Pasqal, Planqc, and QuEra. All authors would like to thank QUTAC.

%%  After submission of the .TEX file,                     %%
%%  you will be prompted to submit your .BBL file.         %%

\bibliographystyle{bmc-mathphys} 
\bibliography{literature}

%%%%%%%%%%%%%%%%%%%%%%%%%%%%%%%%%%%
%%                               %%
%% Figures                       %%
%%                               %%
%% NB: this is for captions and  %%
%% Titles. All graphics must be  %%
%% submitted separately and NOT  %%
%% included in the Tex document  %%
%%                               %%
%%%%%%%%%%%%%%%%%%%%%%%%%%%%%%%%%%%

%%
%% Do not use \listoffigures as most will included as separate files

%%%%%%%%%%%%%%%%%%%%%%%%%%%%%%%%%%%
%%                               %%
%% Additional Files              %%
%%                               %%
%%%%%%%%%%%%%%%%%%%%%%%%%%%%%%%%%%%

%\section*{Additional Files}
%  \subsection*{Additional file 1 --- Sample additional file title}
%    Additional file descriptions text (including details of how to
%    view the file, if it is in a non-standard format or the file extension).  This might
%    refer to a multi-page table or a figure.

%  \subsection*{Additional file 2 --- Sample additional file title}
%    Additional file descriptions text.
%\end{comment}
\end{backmatter}

\end{document}